\documentclass[pra,showpacs,apa,twocolumn,nofootinbib,nobibnotes,superscriptaddress]{revtex4-1}

\usepackage{amssymb}
\usepackage{graphicx,amsmath}
\usepackage{bm}
\usepackage{times}
\usepackage{epstopdf}

\def\expect#1{\left\langle \!#1 \!\right\rangle}
\def\braAket#1#2#3{\left\langle \!#1\! \left|\! #2\! \right|\! #3\! \right \rangle}

\begin{document}

\title{Sources of heading error in optically pumped magnetometers operated at Earth's magnetic field}
\date{\today }
\author{G. Oelsner}
\email{gregor.oelsner@leibniz-ipht.de}
\affiliation{Leibniz Institute of Photonic Technology, P.O. Box 100239, D-07702 Jena, Germany}
\author{V. Schultze}
\affiliation{Leibniz Institute of Photonic Technology, P.O. Box 100239, D-07702 Jena, Germany}
\author{R. IJsselsteijn}
\affiliation{Leibniz Institute of Photonic Technology, P.O. Box 100239, D-07702 Jena, Germany}
\affiliation{Supracon AG, An der Lehmgrube 11, D-07751 Jena, Germany}
\author{F. Wittk\"amper}
\affiliation{Leibniz Institute of Photonic Technology, P.O. Box 100239, D-07702 Jena, Germany}
\author{R. Stolz}
\affiliation{Leibniz Institute of Photonic Technology, P.O. Box 100239, D-07702 Jena, Germany}

\begin{abstract}
When optically pumped magnetometers are aimed for the use in Earth's magnetic field, the orientation of the sensor to the field direction is of special importance to achieve accurate measurement result. Measurement errors and inaccuracies related to the heading of the sensor can be an even more severe problem in the case of special operational configurations, such as for example the use of strong off-resonant pumping. We systematically study the main contributions to the heading error in systems that promise high magnetic field resolutions at Earth's magnetic field strengths, namely the non-linear Zeeman splitting and the orientation dependent light shift. The good correspondence of our theoretical analysis to experimental data demonstrates that both of these effects are related to a heading dependent modification of the interaction between the laser light and the dipole moment of the atoms. Also, our results promise a compensation of both effects using a combination of clockwise and counter clockwise circular polarization.
\end{abstract}

\maketitle

\section{Introduction}
Optically pumped magnetometers (OPM) exploit the natural conversion of magnetic fields to frequencies by the Zeeman effect in alkali atoms \cite{Budker2013}. Because frequencies can be determined with very high precision, magnetometers based on this effect offer the potential of very high sensitivities.

The development of OPMs over the last decades resulted in increased sensitivities that made them competitive with well- established magnetometers based on superconducting systems such as SQUIDs \cite{Allred2002,Kitching2008,Knappe2010,Bevilacqua2013,Lembke2014}. Especially the so-called spin exchange relaxation-free (SERF) type \cite{Allred2002,Happer1973,Kominis2003} is nowadays increasingly applied for the detection of biomagnetic signals \cite{Alem2015,Borna2017}. Additionally, due to their working principle OPMs offer the advantage to be used as total magnetic field sensors \cite{Budker2013,Weis2016} which may be beneficial for a number of applications. Still, a vectorial characteristic is introduced because the laser light photons carry a certain angular momentum in parallel to the beam direction for the purpose of optical pumping spin orientation. Thus, the laser beam direction relative to the magnetic field can falsify the measurement result and influence sensitivity \cite{BenKish2010,Ingleby2017a}. This so-called heading error \cite{Alexandrov2003,Budker2013} leads to certain requirements for the operation of OPMs when used for geomagnetic surveys. For example, their orientation concerning the Earth's magnetic field has to be fixed during the measurement to avoid measurement errors\cite{Stuart1972,Nabighian2010}. Therefore, the study of orientation effects is an important step in the development of new detection concepts with atomic magnetometers and attracts increasing scientific interest \cite{Jensen2009,Colombo2016,Ingleby2017b,Bao2018}.

Much application-driven research on OPMs is orientated to biomagnetic fields. Therein the SERF operational regime allows for femtotesla sensitivities. However, SERF requires an environmental field close to zero (well below 10~nT) \cite{Kominis2003} and sets strong requirements to magnetic shielding \cite{Xia2006} as well as to the compensation of residual fields \cite{Boto2018}. This limitation excludes the SERF-regime for the use in magnetically unshielded environments.

To overcome this obstacle, recently, the light narrowing (LN) \cite{Scholtes2011,*Scholtes2011e} and light-shift dispersed $M_z$ (LSD-Mz) \cite{Schultze2017} operational modes have been developed. Both exploit a strong off-resonant pumping of buffer gas cells \cite{Woetzel2011} for an optimal redistribution of population. This approach results in large output signal amplitudes. Also, the rather broad magnetic resonances, of about 1~kHz, allow for a comparable large bandwidth. In this paper, we analyze the heading error connected to the unusual parameters required for these operational modes. The rest of the paper is organized as follows: In Sec.~\ref{sec:theor} we start our discussion with a description of the atom-light interaction as basis for understanding experimental findings. In Sec.~\ref{sec:exp} the experimental setup and its characterization are demonstrated. We present our experimental findings together with corresponding calculations to the non-linear Zeeman effect and the light shift in respective sections ~\ref{sec:nonlinzeeman} and \ref{sec:lightshift}. Finally, we summarize the paper with a conclusion.

\section{Basic considerations on the atom-light interaction}
\label{sec:theor}
The term heading error summarizes several effects associated with the change of orientation between the laser beam and the external magnetic field $\vec{B}_0$ in OPMs \cite{Budker2013}. In general, two main phenomena can be associated with the LN \cite{Scholtes2011,*Scholtes2011e} and LSD-Mz \cite{Schultze2017} operational modes. First, a significant influence of the non-linear Zeeman splitting \cite{Seltzer2007,Bao2018} can be expected because the suggested magnetometer is aimed for the operation in ambient fields of the order of the Earth's magnetic field. Secondly, the operation in these special modes is based on a strong off-resonant pumping at the optical transitions. This is connected to large light-shifts. Although attempts have been carried out to cancel the effect of non-linear Zeeman splitting by appropriate light-shift \cite{Jensen2009,Chalupczak2010}, such an approach cannot easily be applied to the LN and LSD-Mz mode because large pump-laser powers are required to achieve high field resolutions.

The orientation dependence of both, the non-linear Zeeman splitting and light shift, can be understood by simple considerations: The direction of the magnetic field $\vec{B}_0$ defines the quantization axis in the system that we label as z-axis. For alkali atoms that are usually used for optical magnetometers, the interaction of the $\vec{B}_0$ with the atomic angular momentum $\vec{F}$ as sum of the total electron angular momentum $\vec{J}$ and the nuclear spin $\vec{I}$ leads to the Zeeman splitting, i.e. the removal of the energetic degeneracy of the hyperfine levels. If the induced splitting is small compared to the hyperfine level splitting, meaning small magnetic fields, $F$ is a good quantum number and the sub-levels can be labeled with different quantum numbers $m_F$. Here, the latter is the quantized projection of the atomic angular momentum on the quantization axis.

\begin{figure}[htb]
  \includegraphics[width = 6.8 cm]{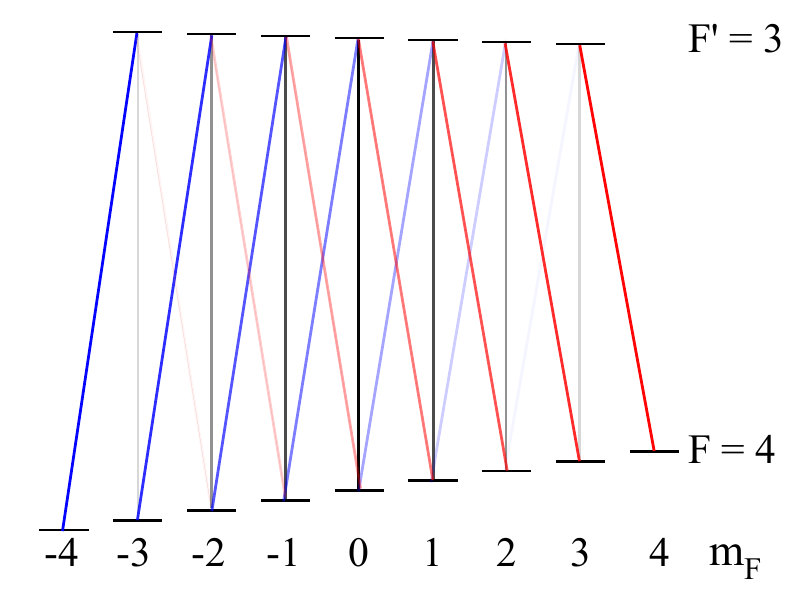}\\
  \caption{(Color online) Transition elements of the $F=4$ to $F^\prime=3$ transition of Cs. The different colors, black, blue, and red are used for visual separation of the $\Delta m_F = 0, \pm 1$ transitions, respectively. The color saturation indicates the strength of the specific transition given by Eq.~\eqref{Eq:transstrength}.}\label{Fig:levpump}
\end{figure}

The directional dependence of OPMs attributes to the interaction of the electric field of the laser beam $\vec{E}(\vec{r},t)$ with the dipole operator $\vec{D}$ of the atom. The electric field is described by a plane wave. We assume that its propagation direction is modified in the x-z-plane. Considering circular polarization, as commonly used for optical pumping, the electric field at the atom's position $\vec{r}_0$ is given by
\begin{equation}
\vec{E}_{\sigma_\pm} =  \frac{E_0}{\sqrt{2}} e^{i\left( \vec{k} \vec{r}_0 - \omega_L t\right)}\left( \cos \alpha \vec{e}_x \pm i \vec{e}_y+ \sin \alpha \vec{e}_z \right).
\end{equation}
Here, $\vec{k}$ and $\omega_L$ are the respective wave vector and angular frequency of the laser light, $\vec{e}_{x,y,z}$ the respective unit vectors in $x,y$, and $z$ direction, and $\alpha$ the angle between the light propagation direction and the z-axis. The plus and minus sign applies for respective left and right handed circular polarization. In the following, we refer to them as $\sigma_+$ and $\sigma_-$ polarization, respectively.

Because the dipole operator is a vector, the matrix elements for optical transitions can be found following the Wigner-Eckart theorem \cite{CohenTan1978}. Introducing $D_\pm = D_x \pm i D_y$, the possible matrix elements for the $F \leftrightarrow F^\prime$ transition of the Cs-$D_1$ line are
\begin{equation}\label{Eq:dipolemoment}
\begin{split}
\braAket{Fm_F}{D_\pm}{F^\prime m_{F^\prime}}&= c_{F,m_F;F^\prime,m_{F^\prime}}^{(\pm1)} \braAket{J}{\left|\vec{D}\right|}{J^\prime}, \\
\braAket{Fm_F}{D_z}{F^\prime m_{F^\prime}} &= c_{F,m_F;F^\prime,m_{F^\prime}}^{(0)}  \braAket{J}{\left|\vec{D}\right|}{J^\prime}.
\end{split}
\end{equation}
The constants $c$ depend on the quantum numbers $F$,$F^\prime$, $m_F$, and $m_{F^\prime}$ and are found by reducing the dipole operator \cite{Steck2010} in terms of Wigner 3-j and 6-j symbols
\begin{equation}\label{Eq:transstrength}
\begin{split}
c_{F,m_F;F^\prime,m_{F^\prime}}^{(q)} \!=&\! \left(-1\right)^{2F^\prime+m_F} \!\sqrt{2\left(2F+1\right)\left(2F^\prime +1\right)} \\
& \left( \!\begin{array}{ccc}F^\prime & 1 & F \\ m_{F^\prime} & q & -m_F \\ \end{array} \!\right) \left\{ \! \begin{array}{ccc}1/2 & 1/2 & 1 \\ F^\prime & F & 7/2 \\ \end{array} \!\right\}
\end{split}
\end{equation}
The above equations represent the optical selection rules $\Delta m_F = 0$ and $q=0$ for linearly as well as $\Delta m_F = \pm 1$ and $q=\pm 1$ for circularly polarized light.

Following \eqref{Eq:dipolemoment}, all possible $F=4 \rightarrow F^\prime = 3$ transitions of the cesium $D_1$ line are illustrated in the level diagram of Fig.~\ref{Fig:levpump}. The ground and excited states are coupled by transition elements resulting from the dipole moment. In dipole approximation the selection rules allow transitions with $\Delta m_F = 0, \pm1 $. The difference in the coupling strengths of the different transitions results in a similar dependence for the rates of spontaneous emission. Note, that there is no additional dependence of the spontaneous decay on the orientation of the laser beam to the magnetic field.

On the other hand, as a consequence of the vector character of the electric field and the dipole operator, the interaction strengths are dependent on the relative angle between beam and magnetic field. Their scalar product can be summarized as
\begin{widetext}
\begin{equation}\label{Eq:interact-energ}
\begin{split}
&{\vec{D}\cdot\vec{E}} = \frac{E_0}{2\sqrt{2}} e^{i\left( \vec{k} \vec{r}_0 - \omega_L t\right)} \left( \left[ \cos \alpha \pm 1\right] D_+ + \left[ \cos \alpha \mp 1\right] D_- +2\sin \alpha D_z \right) , \\
& \braAket{Fm_F}{\vec{D}\cdot\vec{E}}{F^\prime m_{F^\prime}} = \hbar \Omega e^{i\left(\vec{k}\vec{r}-\omega t\right)}\left\{ \begin{array}{ll}  c_{F,m_F;F^\prime,m_{F^\prime}}^{(1)}  \left[ \cos \alpha \pm 1\right],  & m_F^\prime = m_F+ 1, \\
c_{F,m_F;F^\prime,m_{F^\prime}}^{(-1)}  \left[ \cos \alpha \mp 1 \right],  & m_F^\prime = m_F- 1,\\
2 c_{F,m_F;F^\prime,m_{F^\prime}}^{(0)}  \sin \alpha, &m_F^\prime = m_F,
\end{array} \right.
\end{split}
\end{equation}
\end{widetext}
where the driving amplitude
\begin{equation}
\Omega = 
\frac{1}{\hbar}\sqrt{\frac{P_L}{4cn\epsilon_0 A}}\left|\braAket{J=1/2}{\left|\vec{D}\right|}{J^\prime=1/2}\right|,
\end{equation}
depends on the transition dipole matrix element of the $D_1$ transition, the applied laser power $P_L$, and its spot size $A$. Also the speed of light $c$, the permittivity $\epsilon_0$, and the refractive index $n$ were introduced and the double bars denote a reduced matrix element \cite{Steck2010}. As a consequence of Eq.~\eqref{Eq:interact-energ}, all the transitions illustrated in Fig.~\ref{Fig:levpump} are relevant for optical pumping of a magnetometer that is rotated in a magnetic field $\vec{B}_0$. Equation \eqref{Eq:interact-energ} describes a $\sin \alpha$ dependence of the interaction strength for an effective $\pi$ transition ($m_F^\prime = m_F$). Such pumping with linearly polarized light is therefore maximal at perpendicular orientation of the laser beam to the magnetic field $\vec{B}_0$. On the other hand, for angles $\alpha$ away from $90^\circ{}$, the ground state levels $m_F$ are coupled to excited states $m_{F^\prime} = m_F \pm 1$. The respective strengths of each transition is a function of $\alpha$ and optical pumping to spin orientation can be achieved.

\section{Experimental method}
\label{sec:exp}
\subsection*{Measurement setup}
\begin{figure}[htb]
  \includegraphics{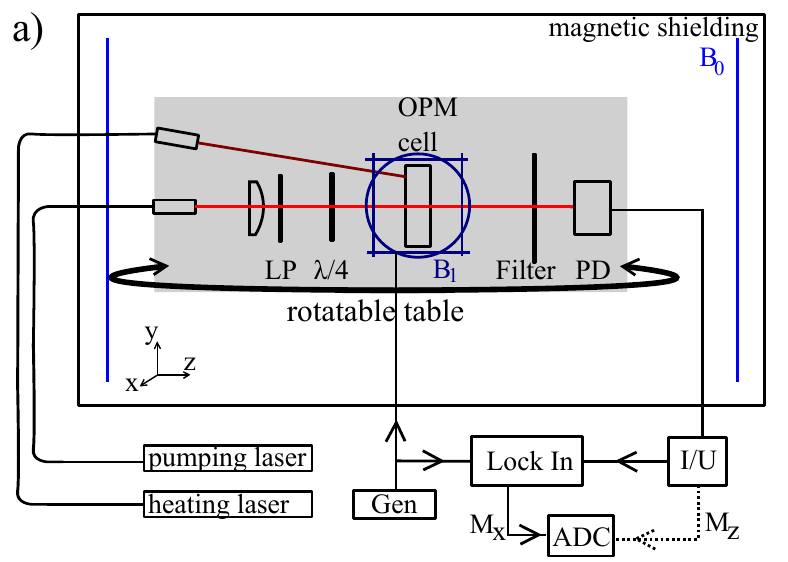}\\
  \includegraphics{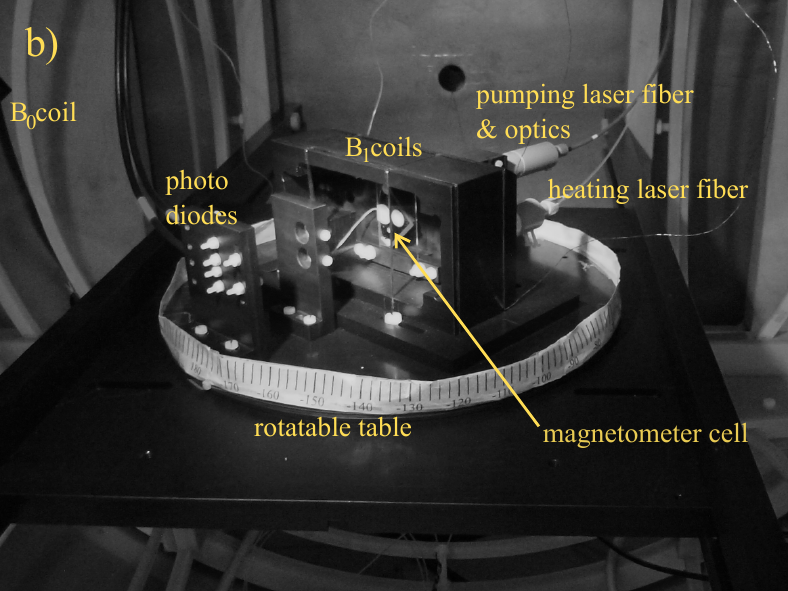}
  \caption{(Color online) Schematic drawing a) and photographic image b) of the measurement setup. A detailed description of the setup is given in the main text. Dependent on the desired experiment a vacuum glass cell and a micro-fabricated buffer gas cell are used, respectively.  }\label{Fig:setup}
\end{figure}

For an experimental investigation of effects associated with the above described modified atom-light coupling, we created a special setup as shown in a schematic drawing accompanied with a photographic image in Fig.~\ref{Fig:setup}. Laser light for pumping the cesium atoms' $D_1$ transition at a wavelength of 895~nm is supplied to the setup by a polarization maintaining fiber. By a combination of a lens, a linear polarizer, and a quarter-wave plate, the pump laser light is collimated and circular polarization is created. The beam diameter is adjusted to 4~mm. This beam is passed either through a vacuum glass cell with paraffin coating \cite{Castagna2009} or a micro-fabricated magnetometer cell \cite{Woetzel2011} both filled with Cs. The latter additionally features a rather high buffer gas pressure of 200~mbar as necessary for both, LN and LSD-Mz operational modes \cite{Scholtes2011,*Scholtes2011e, Schultze2017}. The transmitted laser beam power is detected by a photodiode behind the cell. Ad addtional bandpass filter avoids unwanted light on the diode.

The micro-fabricated magnetometer cell is heated by an additional heating laser beam at a wavelength of 978~nm applied perpendicular to the pump laser beam at the side of the cell. Three coils are mounted perpendicular to each other around the cell allowing applied magnetic $B_1$-fields in all spatial directions.

The described setup is placed on top of a rotatable table inside of a three-axis Helmholtz coil system and three layers of $\ mu$-metal shielding. By that the influence of external magnetic fields is reduced and arbitrary magnetic fields $\vec{B}_0$ can be applied. In the experiment, the direction of $\vec{B}_0$ is kept in z-direction and its strength is set to either 5 or 50~$\mu$T. Note, to avoid unwanted stray fields and to allow a more realistic experiment, in contrast to Refs.~\cite{Scholtes2012,Ingleby2017} we decided to rotate the magnetometer maintaining a constant magnetic field direction. Our setup is adjusted in a way to ensure that the magnetometer cell stays exactly in the center of the $\vec{B}_0$ coil setup during rotation. In this way, the whole setup reflects an atomic magnetometer rotated in an external magnetic field. The alignment of the magnetometer setup with respect to the magnetic field can be controlled from outside the shielding by a cable pull. This setup is somehow comparable to one presented earlier, where the OPM is rotated in the Earth's magnetic field \cite{Hovde2010}, but it excludes external interferences and is variable in the magnetic field strength.

The current measured by the photodiode is amplified by a trans-impedance amplifier before it is demodulated to its in-phase X and quadrature Y component by a lock-in amplifier or directly digitized respectively for $M_x$ or $M_z$ magnetometer operation (see below). A radio frequency generator (Gen) serves as local oscillator for the demodulation and is used to drive one of the $B_1$ field coils.

\subsection*{Measured signals}

To obtain a magnetic resonance signal, the $B_1$ field is applied perpendicular to the direction of $\vec{B}_0$. As often demonstrated \cite{Dehmelt1957,Budker2013,Grosz2016}, the dynamics of optically detected rf-resonances in alkali atoms are well described by the Bloch equation \cite{Bloch1946}. They might also be reconstructed from the Hamiltonian of a two level system
\begin{equation} \label{Eq:Hamiltonian}
H = \hbar \omega_0 \sigma_z + \hbar \Omega_{rf} \cos \omega_{rf} \sigma_x
\end{equation}
in a rotating frame \cite{Oelsner2017a} together with relaxation and decoherence introduced by a Lindblad operator. The parameters in the above equation are the Larmor frequency $\omega_0 = g B_0$, with $g$ being the gyro-magnetic ratio, and the frequency of the rf-field $\omega_{rf}$. In addition, $\hbar \Omega_{rf}$ is the amplitude of the oscillating field, with $\hbar$ the reduced Planck constant. Note, a general $\vec{B}_1$ might be decomposed into a parallel and a perpendicular component with respect to the $\vec{B}_0$ field. The parallel $B_1^{||}$ leads to a modulation of the Larmor frequency. This part can be neglected in a rotating frame. On the other hand, the transition amplitude is given by the perpendicular $B_1^\perp$ field as $\Omega_{rf} = g B_1^\perp$. The relaxation and decoherence rates are named as $\Gamma_r$ and $\Gamma_\varphi$ respectively and correspond to the inverse of the $T_1$ and $T_2$ time.
The steady state solutions for the expectation values of the Pauli operators $\sigma_x$, $\sigma_y$, and $\sigma_z$ in the laboratory frame are given by
\begin{equation}\label{Eq:Blochresults}
\begin{split}
\expect{\sigma_x} &= \frac{\delta \Omega_{rf}}{\Gamma_\varphi \Gamma_\varphi^\prime} \cos \omega t \expect{\sigma_z} + \frac{\Omega_{rf}}{\Gamma_\varphi^\prime} \sin\omega t \expect{\sigma_z} \\
\expect{\sigma_y} &= -i\frac{\delta \Omega_{rf}}{\Gamma_\varphi \Gamma_\varphi^\prime} \sin \omega t \expect{\sigma_z} + i \frac{\Omega_{rf}}{\Gamma_\varphi^\prime} \cos\omega t \expect{\sigma_z} \\
\expect{\sigma_z} &= -\frac{\Gamma_r \Gamma_\varphi^\prime}{\Gamma_r \Gamma_\varphi^\prime + \Omega_{rf}^2}.
\end{split}
\end{equation}
Here $\delta = \omega_0 - \omega_{rf}$ is the detuning between the rf-field and Larmor frequency and $\Gamma_\varphi^\prime = (\Gamma_\varphi^2+\delta^2)/ \Gamma_\varphi$. Depending on the detection technique, either the unmodulated population $\propto \expect{\sigma_z}$ or the off-diagonal components that are modulated at the $B_1$ field frequency can be reconstructed. These methods correspond respectively to the so-called $M_z$ and $M_x$ operational modes of OPMs because the different components in Eq.~\eqref{Eq:Blochresults} are connected to magnetization along different axis (given by the index of the Pauli operators). These components of the magnetization are translated to the absorption of the pump beam and thus can be reconstructed from the measured photo current. An orientation of the laser beam along the $x$ and $z$ axis respectively maximizes the modulated and unmodulated signal because of similar arguments as already presented in Sec.~\ref{sec:theor}. Note, because the pumping of spin orientation breaks down at perpendicular polarization, usually an angle of $45^\circ{}$ is used in the $M_x$ magnetometer mode when using a single laser beam.

\subsection*{Characterization of experimental setup}

A first measurement series is aimed to the characterization of the measurement setup. We use the paraffin coated vacuum cell and a $B_0$ field of $5$~$\mu$T in the $M_x$ regime. The influence of the non-linear Zeeman splitting is reduced by the choice of a comparatively small magnetic field. In addition, the pump laser is adjusted to the $F=4 \rightarrow F^\prime = 3$ cesium $D_1$ transition; and low powers of about 10~$\mu$W reduce the influence of light shifts. All measurements are carried out in $10^\circ{}$ steps over a full rotation of the setup inside the magnetic field. To evaluate the influence of slow magnetic field drifts, the heading angle is first rotated counter clockwise starting at $180^\circ{}$ and measuring in $20^\circ{}$ steps. The intermediate points are then recorded by consequent clockwise rotation.

\begin{figure}[htb]
  \includegraphics{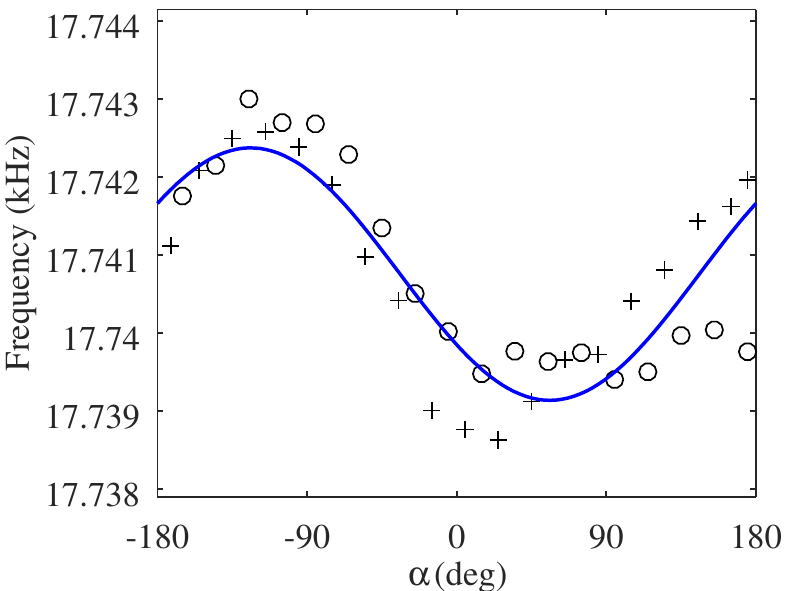}\\
  \caption{(Color online) Measured Larmor frequency as a function of the angle $\alpha$ between the pump laser beam direction and the magnetic field $\vec{B}_0$ measured in the $M_x$ regime on a vacuum cell. The magnetic field was set to about 5~$\mu$T. The crosses and circles correspond respectively to the measurements from 180 to -180 degree and back that are carried out subsequently. The blue line corresponds to a sinusoidal curve fitted to the data with an amplitude of 1.6~Hz. This corresponds to a magnetic field of 0.46 nT, what is about the same magnitude compared to other setups \cite{Hovde2010}. }\label{Fig:vacuum5muT}
\end{figure}

We determine the Larmor frequency in the $M_x$ mode from the in- and out-of-phase signals by a linear fitting of the ratio $X/Y$. Evaluating Eq.~\eqref{Eq:Blochresults} the total population difference $\expect{\sigma_z}$ is canceled
\begin{equation}
X/Y = \frac{\omega_0-\omega_{rf}}{\Gamma_\varphi}
\end{equation}
and the Larmor frequency $\omega_0/2\pi$ as well as the resonance width $\Gamma_\varphi/2\pi$ remain as fitting parameters. In Fig.~\ref{Fig:vacuum5muT} the measured Larmor frequencies are plotted as a function of the orientation $\alpha$ of the laser beam direction to the $B_0$ field vector.

Experimentally we observe a sinusoidal dependence of the Larmor frequency with the orientation. The amplitude of the effect has a value of 1.6~Hz. Despite of a careful experimental setup design, we assume a source of a parasitic field located on the rotatable table that is added and subtracted to the $B_0$ field at different angles and leads to this dependence . Also, while the back and forward measurements (marked by circles and crosses, respectively) give almost the same results close to $-180^\circ{}$, a large deviation of about 2~Hz is found between the start and end point at $+180^\circ{}$. We account this to a low frequency drift of the $B_0$ field that is likely caused by thermal variations of the current sources for the coils or the Helmholtz system itself or the magnetic shielding. This observed remaining orientation dependence is limiting the evaluation of the heading error.

\section{The non-linear Zeeman splitting}
\label{sec:nonlinzeeman}

In the follow-up experiments the $B_0$ field is increased to 50~$\mu$T while the vacuum cell and all other experimental parameters are kept. This setup allows the determination of the heading error caused by the non-linear Zeeman splitting. That is the relative shift of transition frequencies between neighboring ground states depending on the magnetic quantum number $m_F$. It is observable already at Earth's magnetic field strengths. This difference in the splitting of magnetic sub-levels is described by the Breit-Rabi equation \cite{Breit1931}. The equation can be expanded for small magnetic fields, giving in first order the linear Zeeman effect. The corrections found in second order describe the non-linear Zeeman effect important at geomagnetic field strengths sometimes also labeled as quadratic Zeeman effect.

In experiments rotating an optically pumped magnetometer inside a magnetic field, the non-linear Zeeman splitting is observable as broadening of resonance lines and center frequency shift\cite{Jensen2009,Chalupczak2010,Bao2018}. Both these effects are associated to modifications of the ground-states population leading to an overlay of several magnetic resonances in measurements.

\begin{figure}[htb]
  \includegraphics{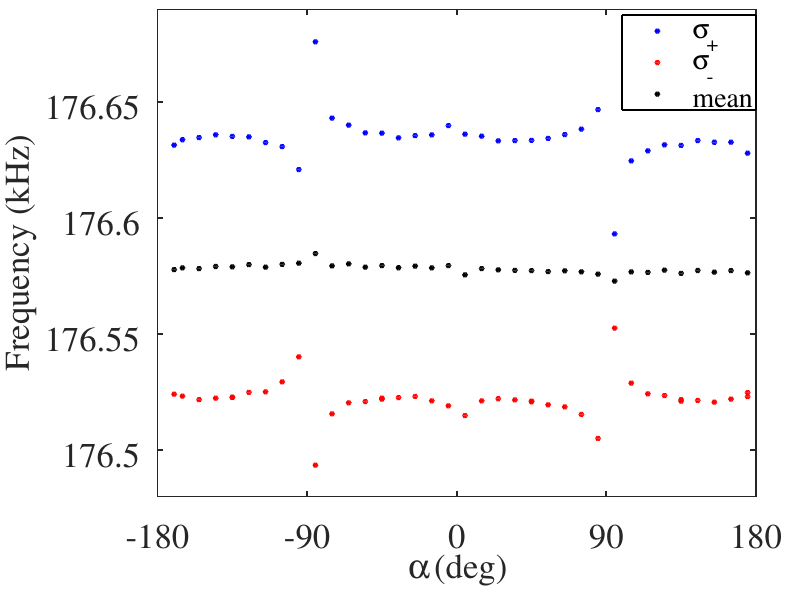}\\
  \caption{(Color online) Measured Larmor frequency as a function of the angle $\alpha$ between the pump laser direction and the magnetic field $\vec{B}_0$ measured in the $M_x$ regime on a paraffin coated vacuum cell. The magnetic field was set to about 50~$\mu$T.}\label{Fig:vacuum50muT}
\end{figure}

In our experimental investigations, we used both left- and right-handed circular polarizations. Thus, the experimental setup had to be modified between the measurement series. Namely the orientation of the quarter wave plate had to be adjusted by opening the mu-metal shielding resulting in a magnetic offset field. Therefore a constant deviation between the Larmor frequencies of both circular polarizations is expected accounting for the magnetic field change due to the rearrangement of the setup. Experimentally we found a value of about 100~Hz. The observed Larmor frequency as a function of the orientation angle is shown in Fig.~\ref{Fig:vacuum50muT}.

The reconstructed Larmor frequency shows an interesting behavior: For $\sigma_+$ polarization and starting from zero, the Larmor frequency decreases with the angle $\alpha$. Towards $90^\circ{}$ the frequency rises strongly before directly at that angle a jump to lower frequencies occurs. Continuing to $180^\circ{}$ we observe a similar but inverse behavior. For $\sigma_-$ polarization the curve is inverted as well. Thus, when taking the mean frequency for both helicities only a weak dependence on the angle is observed that corresponds nicely to the calibration curve in Fig.~\ref{Fig:vacuum5muT}.

In aim to describe these experimental results, we use the results from Sec.~\ref{sec:theor} for an estimation of the change in level population within the frame of rate equations. We focus ons the $F=4$ to $F^\prime = 3$ transition as used in the experimental configuration and consider a paraffin coated vacuum magnetometer cell. In frame of Einstein coefficients, the probability for absorption $W$ is related to the square of the corresponding interaction matrix element's norm as
\begin{equation}\label{Eq:transprob}
W_{m_{F^\prime}m_F} = \frac{\pi}{2\hbar^2} \left|\braAket{4m_F}{\vec{D}\cdot\vec{E}}{3m_{F^\prime}}\right|^2 \! \int_0^\infty \! \rho \!\left( \omega\right) \! s\!\left( \omega \right) d\omega.
\end{equation}

Here, the last term accounts for the spectral distribution of the laser light given by $\rho (\omega)$ and the broadened transitions lines $s(\omega)$ \cite{Steck2018}. Each of those distribution functions is normalized and can be deduced from strict quantum mechanical calculation \cite{Haken1981} or by Fermi golden rule type arguments \cite{Eichhorn2014}. There $s\left(\omega\right)$ defines the spectral distribution of light spontaneous emitted by the considered transition. On the other hand, the laser light will have a narrow spectral width, allowing to set $\rho(\omega) = \delta\left( \omega - \omega_L \right)$. Further, we assume a Lorentzian line shape for the optical transition with an in general Doppler-broadened linewidth $\gamma_{m_{F^\prime}m_F}$. Thus, the integral in Eq.~\ref{Eq:transprob} simplifies to the factor $2\gamma_{m_{F^\prime}m_F}/\pi$ revising Eq.~\ref{Eq:transprob} to
\begin{equation}\label{Eq:abs}
W_{m_{F^\prime}m_F}= c_{m_{F^\prime}m_F}(\alpha)\frac{\Omega^2}{\gamma_{m_{F^\prime}m_F}}.
\end{equation}
Here the angle dependence and the exact value of the transition matrix element are summarized in the factor $c_{m_{F^\prime}m_F}$. In addition to the rate of absorption defined by Eq.~\eqref{Eq:abs}, in our model the nine ground and seven excited states are coupled by spontaneous emission given by the rate
\begin{equation}\label{Eq:decay}
A_{m_{F^\prime}m_F}\!= \!\frac{2\omega_{opt}^3}{3\epsilon_0 h c^3}\! \left|\braAket{4m_F}{\vec{D}}{3 m_{F^\prime}}\right|^2\! =\! c_{F,mF;F^\prime,m_F^\prime}^{(m_F^\prime-m_F)} \gamma_s,
\end{equation}
that is a function of the transition dipole moment. The natural line width $\gamma_s$ for cesium was introduced in Eq.\eqref{Eq:decay}. Note, we neglect the stimulated compared to the much larger spontaneous emission. Also, we added an isotropic relaxation from each ground state to all others with rate $\Gamma_r$.

\begin{figure}[htb]
  \includegraphics{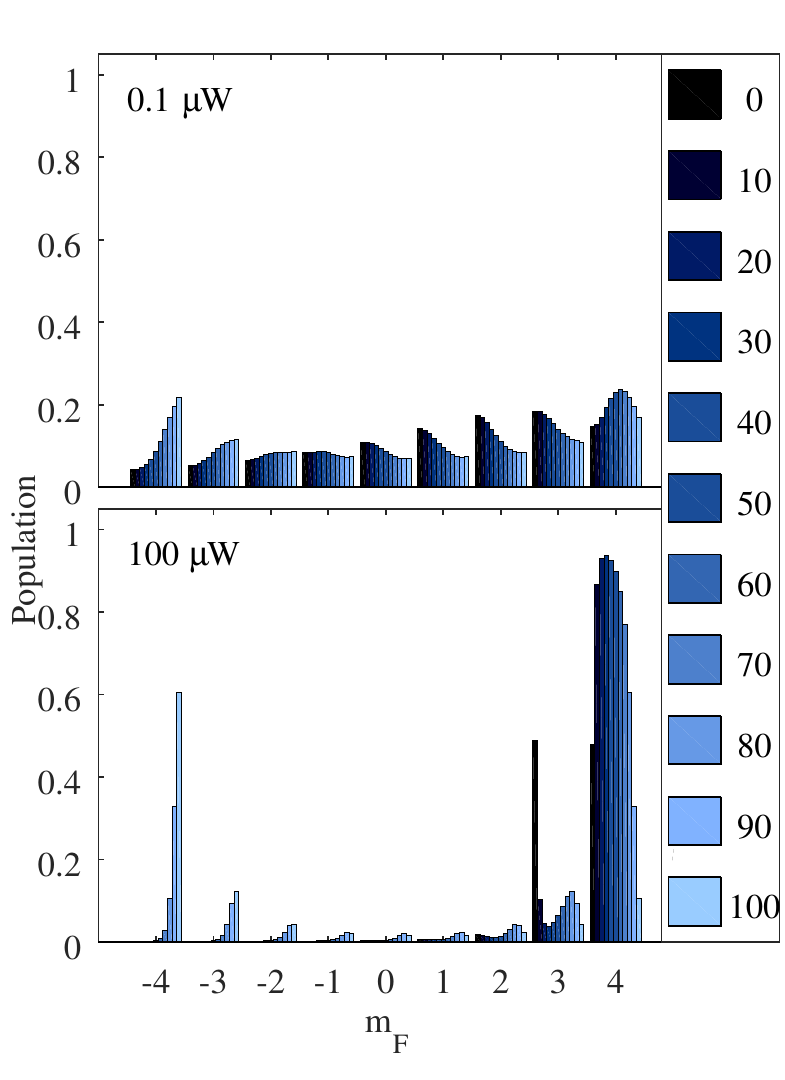}\\
  \caption{(Color online) Population of the $F = 4$ ground state levels estimated by rate equations due to optical pumping with circular polarized radiation $\sigma_+$ at different angles $\alpha$ between magnetic field $\vec{B_0}$ and beam propagation as given by the color. The corresponding numbers at the colormap are given in degree. For the upper and lower picture we assume an integrated laser power of 0.1 and 100 $\mu$W distributed over circular beam shape of 4~mm diameter, respectively. Also a redistribution rate between neighboring ground states of $\Gamma = 25$~Hz is used. The peak at the $m_F=3$ ground state for zero degree in the high pumping case is due to the fact that for optimal circular polarization this level cannot be emptied by the pumping beam. At the other angles a $\pi$-transition component (see Eq.~\ref{Eq:interact-energ}) appears and reduces the population of this state. }\label{Fig:population}
\end{figure}

The discussed model is used to calculate the distribution of population along the ground states numerically. The results for two powers of the pump beam are presented in Fig.~\ref{Fig:population}.

For the simulation, the natural linewidth $\gamma_s/2\pi = 4.5$~MHz of the $D_1$-line corresponding to the rate of spontaneous emission \cite{Steck2010}, a Doppler broadened linewidth of $\gamma_{m_{F^\prime}m_F} = 350$~MHz, and a reduced dipole moment of $\braAket{J=1/2}{\vec{D}}{J^\prime =1/2} = 2.7\times10^{-29}$~Cm were used. Note, only a vanishing population is found in the excited states because the spontaneous emission rate represents the overall fastest process by at least four orders of magnitude.

The the case of low power pumping, we found a rather weak redistribution of the population. Furthermore, for angles close to zero only the blue transitions in Fig.~\ref{Fig:levpump} are of importance and the population is shifted towards the levels with high magnetic quantum numbers $m_F$ as expected when pumping with $\sigma_+$ circular polarization. The strongest population differences are not found for transitions incorporating level with largest $m_F$ but rather around $m_F =3$.

The system is effectively pumped with linearly polarized light when the magnetic field $\vec{B}_0$ is perpendicular to the beam direction. Thus, the population is distributed over many ground state levels leading to much higher absorption of the laser light by the vapor. The blue and red transitions of Fig.~\ref{Fig:levpump} are equally strong ($\cos \alpha = 0$) and the black transitions become maximal in this orientation. By that spin alignment, the concentration of population to states with large absolute quantum numbers $\left|m_F\right|$ \cite{Budker2002,Breschi2012}, is achieved. Simulation results for perpendicular configuration indeed show the largest population differences close to the levels with large absolute magnetic quantum numbers $\left| m_F\right|$.

If the pumping is substantially increased, as shown in the lower plot of Fig.\ref{Fig:population}, a distinct pumping to the dark states is achieved. It is most effective for angles close to zero. If the orientation is changed to $90^\circ{}$, only spin alignment remains. Note, due to the pumping of the $F=4 \rightarrow F^\prime = 3$ transition in principle two dark states are found for optimal $\sigma_+$ pumping. This fact leads to the large population of the $m_F = 3$ state at $\alpha = 0$ and makes pumping to the dark state with $m_F=4$ even more efficient if a small linear polarized component remains in the laser beam. The discussed modifications in the population of states influence the observable magnetic rf-resonance.

The population as plotted in Fig.~\ref{Fig:population} corresponds to the steady state result due to optical pumping. The corresponding rates for optical pumping and spontaneous emission are of the order of tens of kHz and MHz for strong optical pumping, respectively. Thus they are much larger than the dissipative rates $\Gamma_r$ and $\Gamma_\varphi$ of the ground states as well as the driving amplitude $\Omega_{rf}$ of the magnetic resonance that are assumed to be below 10~Hz for a vacuum cell \cite{Castagna2009}. In general, increasing the driving amplitude leads to a significant broadening of the magnetic resonance signal.

However, in the case of strong optical pumping, the exact values of the rates $\Gamma_r$, $\Gamma_\varphi$, and $\Omega_{rf}$ have a marginal influence on the steady-state result and the applied $B_1$ field tends to equalize the population of neighboring magnetic sub-levels. Therefore, the laser absorption needs to be increased depending on the resonance condition between $B_1$ and Larmor frequency to keep the steady state population.

The measurement outcome of a $M_x$ atomic magnetometer rotated in the Earth's magnetic field is estimated by
\begin{equation}\label{Eq:final}
X = \!\sum_{m_F=-4:3}\!-\beta(m_F,\theta)\frac{\Gamma_r \Omega_{rf}}{\Gamma_r \Gamma_\varphi^\prime(\omega_0(m_F)))\! +\! \Omega_{rf}^2}.
\end{equation}
Here, the factor $\beta$ is the population difference of the magnetic transition $m_F \leftrightarrow m_F+1$ and the change of the Larmor frequency due to the non-linear Zeeman effect is included into $\Gamma_\varphi^\prime$, see Eq.~\eqref{Eq:Blochresults}.

\begin{figure}[htb]
  \includegraphics{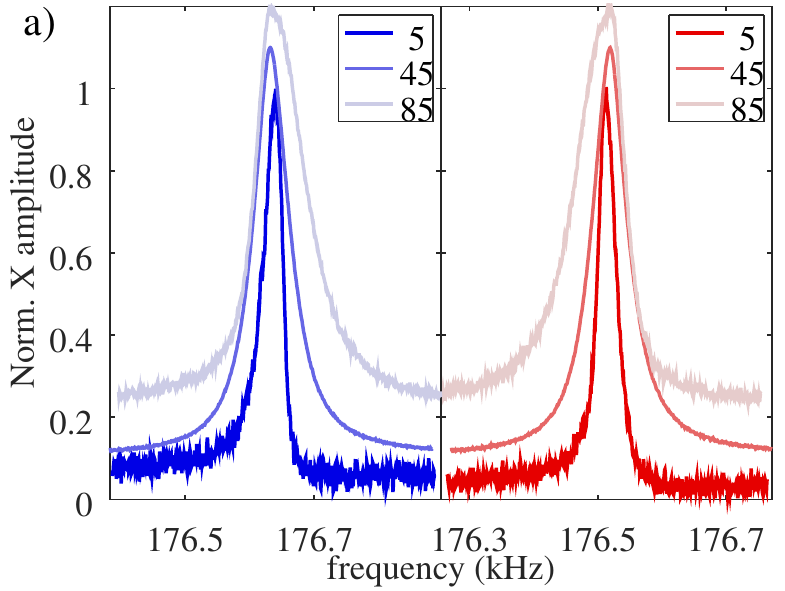}\\
  \includegraphics{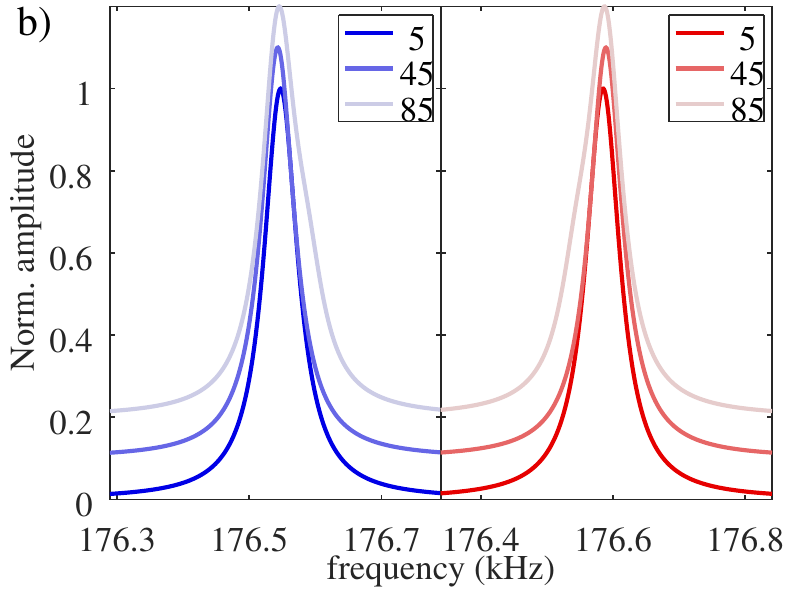}
  \caption{(Color online) a) Example measurement results for the X signal measured at a magnetic field of about 50~$\mu$T on the vacuum cell. The blue and red color are used for $\sigma_+$ and $\sigma_-$ circular polarization of the laser light, respectively. The angles $\alpha$ between the light propagation direction and the magnetic field at which the curves are measured are given by the legend in degree. b) For comparison the results of our simulations are presented in the same style.}\label{Fig:vacuum50muT2}
\end{figure}

Example measurement curves for the X signal are shown together with results of our simulation for an initial test in respective Fig.~\ref{Fig:vacuum50muT2} a) and b). The natural and Doppler broadened linewidths as before (calculation of Fig.~\ref{Fig:population}), transverse as well as longitudinal relaxation times of $T_1= 20$~ms and $T_2 = 10$~ms respectively, and a $B_1$ amplitude $\Omega_{rf}$ of $20$~Hz  were used. The ground state dissipation rates are slightly increased compared to the values as for example given in Ref.~\cite{Castagna2009}, but were required to get a reasonable redistribution of population to fit the experimental results. Also we used a laser power of 50~$\mu$W on a 4~mm diameter spot and a magnetic field $B_0 = 50.34$~$\mu$T, approximately corresponding to the experimental situation.

The experimental curves in Fig.~\ref{Fig:vacuum50muT2}~a) nicely demonstrate that the measured shift of the resonance line is mainly caused by a broadening of the magnetic resonance curve. It indicates a distributed population for certain angles $\alpha$ that results from an overlay of several Lorentzian curves with shifted center frequencies as a result of the non-linear Zeeman splitting. The same process leads to the asymmetry of the curves \cite{Budker2013}. Our model is able to reproduce the experimental results as shown in Fig.~\ref{Fig:vacuum50muT2}~b). Accordingly, we observe the same shift of the resonance curves as well as their broadening towards $90^\circ{}$. Still, especially the linewidth at five degrees is larger than the one found in the experiment. In this context, we note that our model does not cover the spin-exchange relaxation accurately and we might overestimate the laser intensity because of assuming a uniform distribution of power over a 4~mm disc. We conclude that an accurate description of the population's redistribution process in the ground state including spin-exchange (as for example in Ref.~\cite{Shi2018}) as well as the influence of the $B_1$ field is required for a more precise prediction of the line shapes.

\begin{figure}[htb]
  \includegraphics{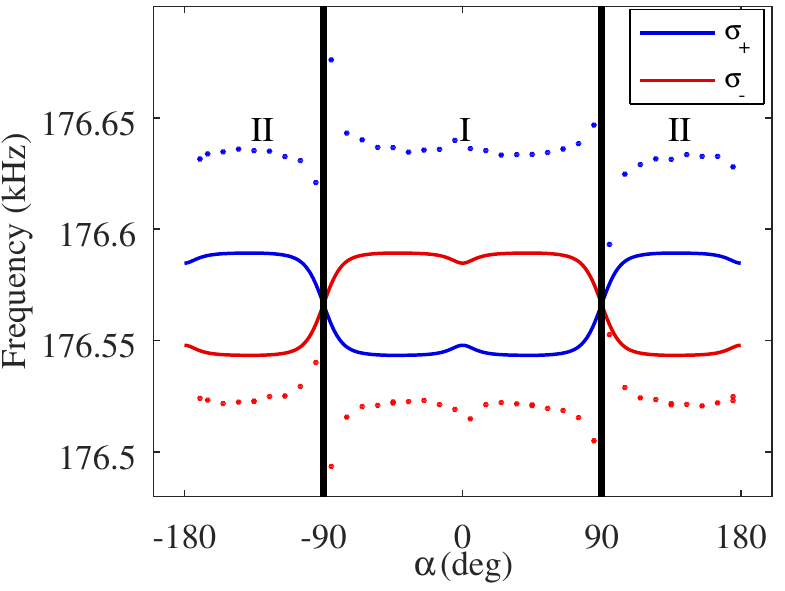} \\
  \caption{(Color online) The Larmor frequency as expected from a measurement as a function of the orientation angle $\alpha$ as straight lines together with the experimental data points added as dots. }\label{Fig:theorlines}
\end{figure}

In Fig.~\ref{Fig:theorlines} we plotted the center frequency of the theoretically predicted line shapes, given by \eqref{Eq:final}, together with the experimental results. The former are found by Lorentzian fits of the theoretical curves exemplarily shown in Fig.~\ref{Fig:vacuum50muT2}. Therewith the plot describes the heading error of the Larmor frequency due to the non-linear Zeeman splitting.

Our simulation perfectly supports our expectations. Particularly, for $\sigma_+$ polarization (blue curve, $\left|\alpha\right|<90^\circ$) we are effectively probing the transitions between the ground states of highest magnetic quantum numbers because the population is concentrated there. If the helicity is changed to $\sigma_-$, either by the orientation of the quarter-wave plate (red curve) or by adjusting the laser antiparallel to the magnetic field (blue curve, $\left|\alpha\right|>90^\circ{}$), we probe effectively the ground states with lowest $m_F$. Due to the non-linear Zeeman splitting the latter has an about 50~Hz larger transition frequency. At angles close to $90^\circ{}$ we find the transition between the $\sigma_+$ and $\sigma_-$ regime that is connected to an increased linewidth of the magnetic resonance as demonstrated in Fig.~\ref{Fig:vacuum50muT2}. The steepness of the crossover depends on the ratio of pumping and ground state redistribution.

The peculiarities at 0 and 180 degrees are connected to the use of the $F=4 \rightarrow F^\prime = 3$ transition for optical pumping. The occurrence of two dark states for perfect circular polarization leads to a roughly equal population of the $m_F = 3$ and $m_F = 4$ ground state (see Fig.~\ref{Fig:population}) and a large population difference between $m_F = 2$ and $m_F = 3$ for $\sigma_+$ at $\alpha = 0$. Thus at this angle a slightly higher center frequency is found for the magnetic resonance compared to larger angles. There the appearing linear pumping component ($\propto \sin \alpha$ in Eq.~\eqref{Eq:interact-energ}) results in the removal of population from the $m_F = 3$ ground state. The same process leads to a reduction of the measured frequency at $0^\circ{}$ for $\sigma_-$ light and also creates the peak of population at $0^\circ{}$ at the $m_F=3$ ground state in the lower plot of Fig.~\ref{Fig:population}. Such effects could be avoided by pumping to the $F^\prime = 4$ excited state.

Interestingly, the shape of our simulated curve corresponds very well to the experimental results, except for some offset frequency. These offset frequencies are about 100~Hz and 50~Hz for the blue curve in the respective marked regions I and II as well as -70~Hz and -20~Hz for the red. While we can explain an offset between the blue and red curve by a field introduced by the required modification of the experimental setup (namely adjusting the quarter-wave plate), the origin of the different offsets in regions I and II is unresolved. With our calibration measurement we can exclude effects from light shift or the setup. Thus, this shifting of the lines has to be connected with the increase in the magnetic field from 5 to 50~$\mu$T. We note that the paramagnetic term of the atoms Hamiltonian is covered by the Breit-Rabi equation. Therefore, the observed shift of the magnetic resonances might be due to the diamagnetic term that scales also with the square of the magnetic field. Essentially, this unresolved effect reduces the experimentally observed heading error to angles close to $\pm 90^\circ{}$.

\section{Light shift in the LN and LSD-Mz operational mode}
\label{sec:lightshift}
For the following measurements the experimental setup as shown in Fig.~\ref{Fig:setup} is extended for the simultaneous recording of both circular polarizations.\footnote{Note, this is possible for our micro-fabricated buffer gas cells, because they offer flat glass surfaces. Still the same cannot easily be done for the glass-blown vacuum cell.} Thus, a polarizing beam splitter and a deflecting prism are used to create two beams and an additional linear polarizer is introduced in the straight beam for intensity adjustment. Both beams are circularly polarized by separate quarter-wave plates and passed through micro-fabricated vapor cells that share a common reservoir. This cell is heated to about 380~K by the 978~nm laser radiation. In experiments with micro-fabricated cells we observe optical linewidths in the order of several GHz due to the high buffer gas pressure \cite{Scholtes2011,*Scholtes2011e}. Therefore, while the ground state hyperfine splitting is resolved the excited states of the cesium $D_1$ transition overlap. The wavelength of the pump laser is stabilized to the $F=3 \rightarrow F^\prime = 4 $ transition on a Doppler-free absorption spectrum of an additional vacuum cell. This modification of the wavelength is required to effectively deplete the $F=3$ ground state levels as required by the LN and LSD-Mz operational modes. Due to slightly overlapping absorption lines, the same laser optically pumps the $F=4$ levels to the dark states $\Delta m_F = \pm 1$ depending on the circular polarization direction. A large light shift of the energy levels is observable in the magnetic transition between the Zeeman split states as expected for strong off-resonant pumping.

\begin{figure}[htb]
  \includegraphics{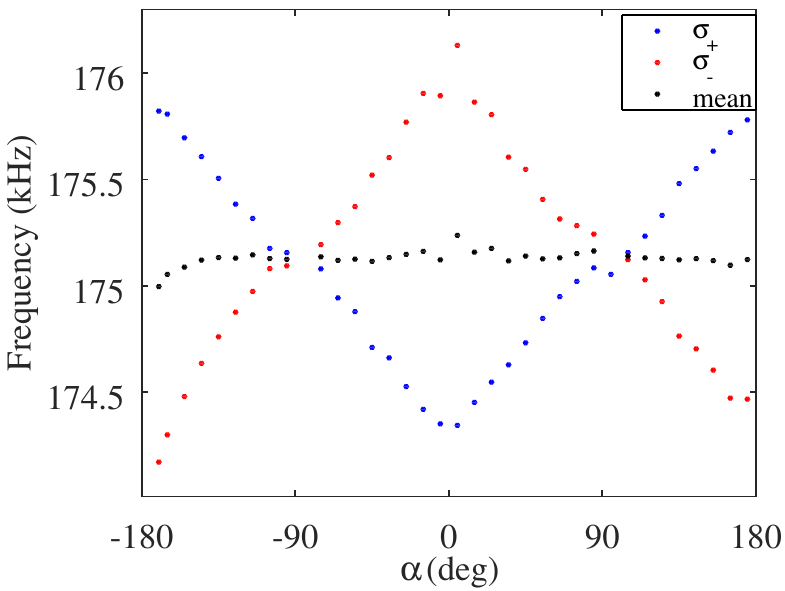}\\
  \caption{(Color online) Measured Larmor frequency as a function of the angle $\alpha$ in degrees between the pump laser direction and the magnetic field $B_0$ measured in the LN regime on a micro fabricated cell at about $B_0=50$~$\mu$T.} \label{Fig:LN}
\end{figure}

\begin{figure}[htb]
  \includegraphics{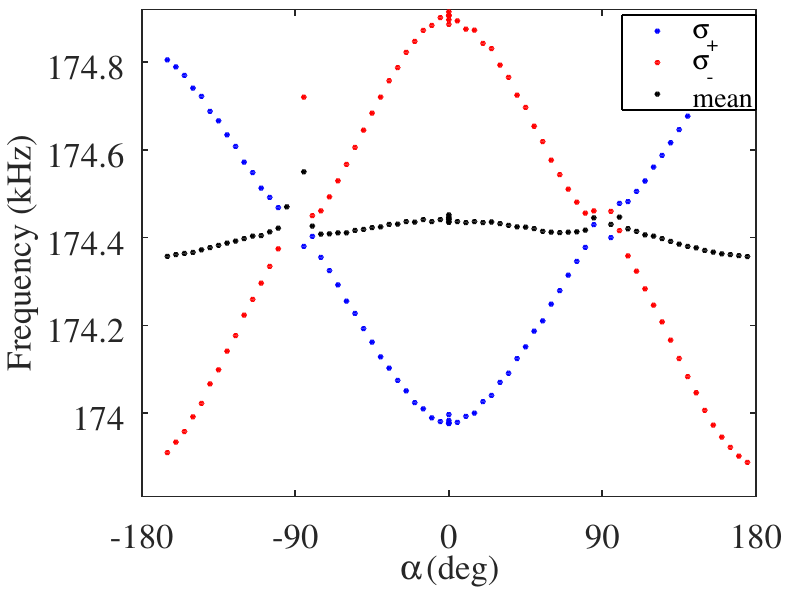}\\
  \caption{(Color online) Measured Larmor frequency as a function of the angle $\alpha$ in degrees between the pump laser direction and the magnetic field $B_0$ measured in the LSD-Mz regime on a micro fabricated cell at about $B_0=50$~$\mu$T.} \label{Fig:LSDMz}
\end{figure}

We present measurements of the orientation dependence of the measured Larmor frequency for the LN and LSD-Mz mode in Figs.~\ref{Fig:LN} and \ref{Fig:LSDMz}, respectively. In both cases a large light shift of the order of 1~kHz for each of the two circular polarizations is observed. Note, optical pumping in both cases works well at 0 and 180 degrees but also the largest light shifts are observed at these angles. Secondly, at these angles the LN-signal is minimal, because the modulated components are observed best in the x-y plane as demonstrated by Eq.~\eqref{Eq:Blochresults}. This is most probably causing the discrepancy of the curve of the measured Larmor frequencies in the LN-$M_x$ method compared to the LSD-Mz mode. The laser power of each beam in the experiment was about 2~mW in the LSD-Mz mode and slightly larger for the LN-regime.

Additionally, we experimentally observe a larger light shift for the $\sigma_-$ polarization compared to that of $\sigma_+$. This discrepancy is most likely caused by an imbalance of laser intensities and/or polarizations of the two beams. Therefore, especially in Fig.~\ref{Fig:LSDMz}, a dependence of the mean Larmor frequency of the two polarization directions on the rotational angle remains.

\begin{figure}[htb]
  \includegraphics{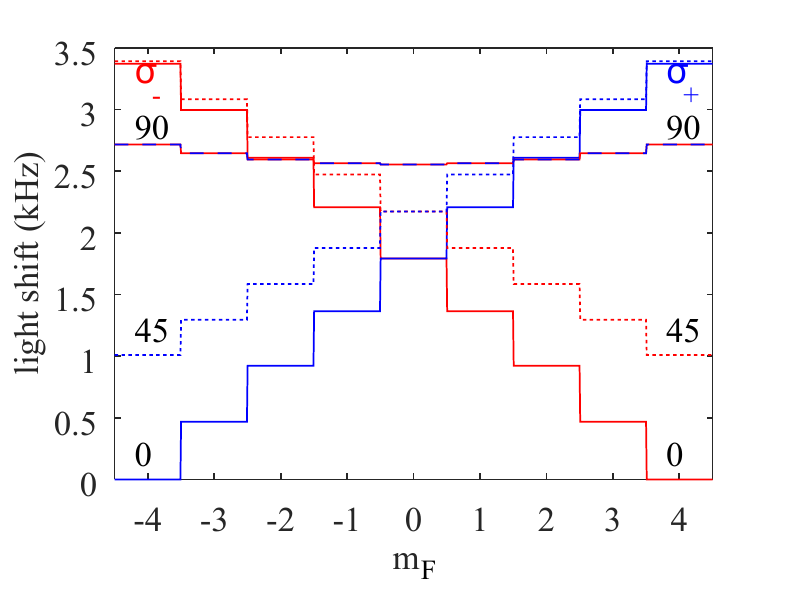}\\
  \caption{(Color online) Light shift of the energy levels for the $F=4$ ground state levels. They are respectively plotted for $\sigma_-$ and $\sigma_+$ polarized light in red and blue. The solid, dotted, and dashed lines correspond to orientation angles of 0, 45, and 90 degrees between magnetic field and laser beam direction. The parameters for the plot are given in the main text. }\label{Fig:lightshift}
\end{figure}

An expression for the light shift in the case of a far from resonance detuned laser is given by \cite{LeKien2013}
\begin{equation}\label{Eq:energyshift}
\begin{split}
\Delta \omega_{m_F} \!=& \frac{-1}{4\hbar^2} \!\sum_{F^\prime m_{F^\prime}}\!\text{Re}\! \left(\!  \frac{\left|\braAket{4m_F}{ \vec{E}\cdot\vec{D}}{F^\prime m_{F^\prime}  } \right|^2  }{\omega_{F^\prime m_{F^\prime}}\!-\!\omega_{4m_F}\!-\!\omega_L\!-\!i\gamma_{F^\prime m_{F^\prime}m_F}/2} \right. \\
& -\left. \frac{\left| \braAket{4 m_F}{ \vec{E}\cdot\vec{D}}{F^\prime m_{F^\prime}}  \right|^2  }{\omega_{F^\prime m_{F^\prime}}-\omega_{4m_F}+\omega_L-i\gamma_{F^\prime m_{F^\prime}m_F}/2} \right) .
\end{split}
\end{equation}
Here, the energy of the i-th state is $\hbar \omega_{i}$. For simplicity the detuning between the transition frequency of the $F=4, m_F$ ground to $F^\prime, m_{F^\prime}$ excited state $\omega_{4m_F,F^\prime m_{F^\prime}}=\omega_{F^\prime m_{F^\prime}}-\omega_{4m_F}$ and the laser frequency $\omega_L/2\pi$ is set constant for all of the transitions $\delta_{4m_F, F^\prime m_{F^\prime}} = \omega_{F^\prime m_{F^\prime}}-\omega_{4m_F}-\omega_L = \delta_{F^\prime}$. This approximation is justified because the latter is significantly larger than the Zeeman splitting of the ground and excited state for the considered operational modes. Further simplifying, one can neglect the second term in the brackets of Eq.~\eqref{Eq:energyshift} since it is significantly smaller than the first. Assuming a constant width of each of the transitions $\gamma$, Eq.~\eqref{Eq:energyshift} takes the compact form
\begin{equation}\label{Eq:energyshift2}
\Delta \omega_{m_F} = \frac{-1}{4\hbar^2} \sum_{F^\prime m_{F^\prime}} \text{Re} \left(  \frac{\left|\braAket{pF^\prime m_{F^\prime}}{ \vec{E}\cdot\vec{D}}{ s4m_F } \right|^2  }{\delta_{F^\prime}-i\gamma/2} \right).
\end{equation}
Above equation can easily be interpreted reminding the light shift of a two level system. The latter is easily reconstructed by expanding the generalized Rabi frequency $\Omega_R$\cite{Oelsner2013} for small interaction energies $\hbar \Omega_0$ compared to the energetic detuning $\hbar \delta$
\begin{equation}\label{Eq:Rabi}
\Omega_R = \sqrt{\Omega_0^2+\delta^2} \approx \delta + \frac{\Omega_0^2}{2 \delta} + O(x^4).
\end{equation}
Note that the frequency shift for one of the levels is $\Omega_R/2$ and the second term describes the power dependent shift of the energy levels. Also these oscillations get damped by a dephasing $\gamma$ leading to a smaller measured splitting as included already in Eq.\ref{Eq:energyshift2}. Therefore this equation describes the sum of all the measurable frequency shifts induced by the coupling of a single ground state level to all allowed excited states. This shift is increased by larger intensities $\propto \vec{E}^2$ or smaller detuning. One should recognize, instead of using the expansion in Eq.~\ref{Eq:Rabi}, the use of the generalized Rabi frequency in principle allows a description for arbitrary detuning, namely in frame of dressed states. The selection rules are included in the matrix elements of $\vec{E}\cdot\vec{D}$. Their modification by the change of orientation of the laser light concerning the $\vec{B}_0$ direction is responsible for the heading dependent light shift and already given in Eq.~\ref{Eq:interact-energ}. In Fig.~\ref{Fig:lightshift} the shifts of the energy $F=4$ ground states are plotted for different angles between magnetic field and laser beam direction.

Here, for each $F=4$ ground state level the sum given by Eq.~\eqref{Eq:energyshift2} is calculated considering all transitions to the $F^\prime = 3,4$ excited states that have non-zero matrix elements \eqref{Eq:interact-energ} in the dipole approximation. The respective detunings of the laser to the transition frequencies are $\delta_3/2\pi = -8.6$~GHz and $\delta_4/2\pi = -9.8$~GHz. Also a transition linewidth $\gamma/2\pi = 4$~GHz and driving amplitude of $\Omega/2\pi = 4$~MHz were used. The latter corresponds to a laser power of 1.3~mW estimated for a perfect circular beam shape on a spot size of 4~mm diameter and thus represents well the LSD-Mz measurement case.

In Fig.~\ref{Fig:lightshift} we observe all the light shift contributions discussed in literature \cite{Hu2018}. All the energy levels are shifted by a constant value. For an orientation of $90^\circ{}$, this shift takes a value of about 2.5~kHz and it is slightly smaller at the other shown angles. It corresponds to the scalar light shift. Secondly, the vector part is most pronounced at $0^\circ{}$, indicating the positions where optimal circularly polarized pumping is achieved. It describes a shift of the energy levels proportional to the magnetic quantum number $m_F$. The direction of this shift is changed between the two different circular polarizations. This vector light shift can be interpreted as additional magnetic field $\vec{B}_{LS}$ added to $\vec{B}_0$ that shifts the Larmor frequencies. Finally, the tensor part is most pronounced at an angle of $90^\circ{}$ and corresponds to a small shift for large absolute magnetic quantum numbers $\left| m_F\right|$.
\begin{figure}[htb]
  \includegraphics{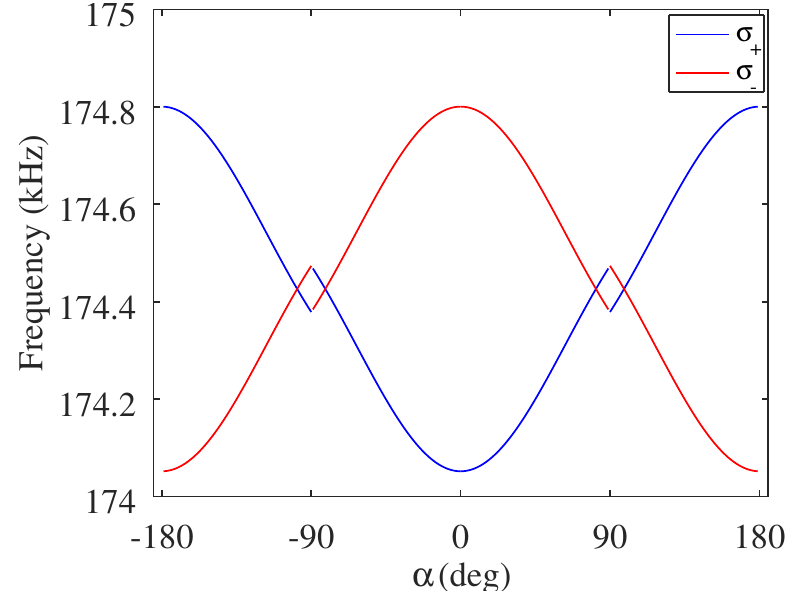}\\
  \caption{(Color online) Expected reconstructed Larmor frequencies as a function of the direction angle $\alpha$. They are influenced by a strong light shift due to the strong off-resonant pumping that is modified due to the change of interaction matrix elements, see Eq.~\ref{Eq:interact-energ}. The same parameters as for Fig.\ref{Fig:lightshift} have been used. }\label{Fig:lightshift2}
\end{figure}

The effect of this shift of the energy levels to the measurable Larmor frequency as a function of the orientation angle $\alpha$ is demonstrated in Fig.~\ref{Fig:lightshift2}. Here we plotted the distance of the relevant dark state to its neighboring level. That is the distance between $m_F =3 \rightarrow 4$ transition for $\sigma_+$ and $\sigma_-$ polarized light in the respective range of [-90 to 90] degrees and [-180 to -90, 90 to 180] degrees as well as the distance between $m_F =-4 \rightarrow -3$ for the same polarizations with interchanged angle ranges. Note this creates two possible frequency values for each polarization at $\pm 90^\circ{}$ which is not realized in the experiment. There the redistribution of population will result in an overlay of several magnetic transitions as discussed for the non-linear Zeeman splitting. Still, we keep the presentation in that way to illustrate the effect of the tensor light shift on the energy levels.

The theoretical plots of Fig.~\ref{Fig:lightshift2} agree qualitatively as well as quantitatively very well with the experimental curves in the LSD-Mz regime of Fig.~\ref{Fig:LSDMz}. Here we used a magnetic field $B_0 = 49.5$~$\mu$T. We observe the same large measurable light shift at angles of $0^\circ{}$ and $180^\circ{}$ connected to a maximal vector light shift due to optimal circularly polarized pumping. It is then reduced reaching values of $\left|90^\circ{}\right|$. An overshooting of the corresponding curves is visible at these angles because the tensor component of the light shift remains. That is also indicated in the experimental curve. In this context, please note that in the experiment the amplitude of the magnetic resonance becomes small at angles of $90^\circ{}$. Also because the linewidth of the magnetic resonance is of the order of 1~kHz in the LSD-Mz mode, the alignment of the magnetic population and therewith the tensor light shift cannot be resolved.

\section*{Conclusion}
We have investigated main contributions to the heading error for operation of optically pumped magnetometers with strong off-resonant pumping at Earth's magnetic field strengths. A dedicated experimental setup is used for a realistic measurement environment, namely a well-defined magnetic field in which the alignment of a magnetometer can be adjusted. The influence of the non-linear Zeeman splitting and of light shifts is evaluated in two separate sets of experiments. Our theoretical model suggests that the first results from a redistribution of the ground states' population. Thus the experimental dependence on the angle is strongly connected to the pump beam intensity and the redistribution process of the ground state population. Also, we observed an unresolved shift in the experiment compared to the theoretical results depending on the orientation of the magnetic field and the polarization of the laser beam. That might indicate a contribution of the induced magnetic dipole in the atomic shell and remains a subject for further investigations.

The light shift has three components: First, the vector light shift that adds a virtual field in the direction of the magnetic field but not in the direction of the pump beam. Secondly, while rotating the magnetometer towards perpendicular orientation according to the magnetic field contributions from tensor light shifts increase. The third is scalar component that gives a constant offset over all angles, and should be considered in the use of a optically pumped magnetometer as absolute field sensor. Having found that the two described contributions to the heading error can be compensated by the use of $\sigma_+$ and $\sigma_-$ light, further work should evaluate the sensitivity of the discussed operational modes.

\section*{Acknowledgments}
G.O. thanks E. Il'ichev for valuable discussions. We like to acknowledge the careful proof reading by T. Scholtes and S.V. Vegesna. The project 2017~FE~9128, funded by the Free State of Thuringia, was co-financed by European Union funds under the European Regional Development Fund (ERDF). This work was conducted using the infrastructure supported by the Free State of Thuringia under the grant number 2015~FGI~0008 and co-financed by European Union funds under the European Regional Development Fund (EFRE).

\bibliography{lit_heading}
\end{document}